\documentclass[twocolumn]{emulateapj}

\usepackage{amsmath}
\usepackage{adjustbox}
\usepackage[title]{appendix}

\shorttitle{An Ultramassive White Dwarf in NGC 2099}
\shortauthors{Cummings et~al.}

\begin{document}

\title{An Ultramassive 1.28 M$_\odot$ White Dwarf in NGC 2099\textsuperscript{1}}


\author{Jeffrey D. Cummings\altaffilmark{2}, Jason S. Kalirai\altaffilmark{3,2},  P.-E. Tremblay\altaffilmark{4}, Enrico Ramirez-Ruiz\altaffilmark{5}, AND P. Bergeron\altaffilmark{6}}
\affil{}

\footnotetext[1]{Based on
observations with the W.M. Keck Observatory, which is operated as a scientific partnership
among the California Institute of Technology, the University of California, and NASA, was made 
possible by the generous financial support of the W.M. Keck Foundation.}

\altaffiltext{2}{Center for Astrophysical Sciences, Johns Hopkins University,
Baltimore, MD 21218, USA; jcummi19@jhu.edu}
\altaffiltext{3}{Space Telescope Science Institute, 3700 San Martin Drive, Baltimore, MD 21218, USA;
jkalirai@stsci.edu}
\altaffiltext{4}{Department of Physics, University of Warwick, Coventry CV4 7AL, UK; 
P-E.Tremblay@warwick.ac.uk} 
\altaffiltext{5}{Department of Astronomy and Astrophysics, University of California,
Santa Cruz, CA 95064; enrico@ucolick.org} 
\altaffiltext{6}{D{\'e}partement de Physique, Universit{\'e} de Montr{\'e}al, C.P. 6128, Succ. 
Centre-Ville, Montr{\'e}al, QC H3C 3J7, Canada; bergeron@ASTRO.UMontreal.CA}

\begin{abstract}
With the Keck I Low-Resolution Imaging Spectrometer we have observed nine white dwarf candidates
in the very rich open cluster NGC 2099 (M37).  The spectroscopy shows seven to be DA white dwarfs, one
to be a DB white dwarf, and one to be a DZ white dwarf.  Three of these DA white dwarfs are consistent
with singly evolved cluster membership: an ultramassive (1.28$^{+0.05}_{-0.08}$ M$_\odot$) 
and two intermediate-mass (0.70 and 0.75 M$_\odot$) white dwarfs.  Analysis of their cooling
ages allows us to calculate their progenitor masses and establish new constraints on the 
initial-final mass relation.  The intermediate-mass white dwarfs are in strong agreement with 
previous work over this mass regime.  The ultramassive white dwarf has $V$ = 24.5, 
$\sim$2~mag fainter than the other two remnants. The spectrum of this star has lower 
quality, so the derived stellar properties (e.g., T$_{\rm eff}$, log g) have 
uncertainties that are several times higher than the brighter counterparts.  We measure these 
uncertainties and establish the star's final mass as the highest-mass white dwarf discovered 
thus far in a cluster, but we are unable to calculate its progenitor mass because at this
high mass and cooler T$_{\rm eff}$ its inferred cooling age is highly sensitive to its mass.
At the highest temperatures, however, this sensitivity of cooling age to an ultramassive 
white dwarf's mass is 
only moderate.  This demonstrates that future investigations of the upper-mass end of the 
initial-final mass relation must identify massive, newly formed white dwarfs (i.e., in young 
clusters with ages 50-150~Myr).
\end{abstract}

\section{Introduction}

White dwarfs that are members of well-studied star clusters are extremely valuable for 
understanding the process of stellar evolution and mass loss.  The progenitor masses (hereafter 
M$_{\rm initial}$) of these white dwarfs can be 
calculated by comparing the remnant's cooling age to the cluster age, a technique that has now 
led to a well established initial-final mass relation (IFMR) from M$_{\rm initial}$ = 0.8 to 5~M$_\odot$ (e.g., 
Claver et~al.\ 2001; Dobbie et~al.\ 2004, 2006a; Williams et~al.\ 2004; Kalirai et~al.\ 
2005; Liebert et~al.\ 2005; Williams \& Bolte 2007; Kalirai et~al.\ 2007; Kalirai et~al.\ 2008; 
Rubin et~al.\ 2008; Kalirai et~al.\ 2009; Williams et~al.\ 2009; Dobbie 
et~al.\ 2012; Cummings et~al.\ 2015, hereafter Paper I; Cummings et~al.\ 2016, hereafter Paper II).  
At higher white dwarf masses (hereafter M$_{\rm final}$), the relation remains poorly 
constrained.  The progenitors of these massive white dwarfs are intermediate-mass stars that 
quickly evolve to asymptotic giant branch (AGB) stars, which lose mass through dust driven 
outflows and thermal pulses.  This phase of stellar evolution is poorly understood from first 
principles and is difficult to model through direct observations.  New constraints on the 
IFMR in this regime would be a breakthrough for stellar astrophysics.

Finding and characterizing high-mass white dwarfs is longstanding challenge due to 
their scarcity.  In the Sloan Digital Sky Survey and Palomar Green Survey only 1.5\%\, and
2.6\%, respectively, of the field white dwarfs have a M$_{\rm final}$$\geq$1.05 M$_\odot$ (e.g., 
Kleinman et~al.\ 2013; Kepler et~al.\ 2016; Liebert et~al.\ 2005).  In star clusters their number 
remains limited at six.  A long known high-mass white dwarf is LB 1497 from the young Pleiades 
star cluster at 1.05 M$_\odot$ (Gianninas et~al.\ 2011).  The remaining five have been recently 
discovered: NGC 2287-4 (Dobbie et~al.\ 2012), NGC 2168-LAWDS27 (Williams et~al.\ 2009), two 
white dwarfs in NGC 2323 (Paper II), and VPHASJ1103-5837 in NGC 3532 (Raddi et~al.\ 2016).  
VPHASJ1103-5837 has a M$_{\rm final}$$\sim$1.13 M$_\odot$ and the four others all have a 
M$_{\rm final}$$\sim$1.07 M$_\odot$ (Paper II).  Two special cases are GD50 at 1.25$\pm$0.02 
M$_\odot$ and PG 0136+251 at 1.19$\pm$0.03 M$_\odot$, which are ultramassive white dwarfs with 
possible connection to the Pleiades.  Based on GD50's space motion, Dobbie et~al.\ (2006b) find 
a high probability it is coeval with the young Pleiades and that it was ejected from the cluster.  
Similar analysis of PG 0136+251 finds provisional connections to 
the Pleiades based on its proper motion, but its radial velocity is still needed to verify
this connection (Dobbie et~al.\ 2006b).  The scarcity of massive white dwarfs in the Galactic 
field, but even more so in stellar clusters, has led to arguments that most massive white 
dwarfs are formed through mass transfer or white dwarf mergers, which theoretically can form such 
massive white dwarfs (e.g., Dan et~al.\ 2014).  These merger processes may create excess massive 
white dwarfs in the field but would not yet play a significant role in the younger cluster populations.  

There are several reasons, however, that can explain this scarcity besides the challenge that their progenitors 
(M$_{\rm initial}$ $>$ 6 M$_\odot$) are rare.  These include that: (1) Increasingly higher-mass
white dwarfs become more compact under their strong gravities, which gives them significantly smaller 
radii and luminosity in comparison to their lower-mass companions.  (2) These white dwarfs form from 
rapidly evolving higher-mass stars, which means that in most clusters they have already 
undergone significant cooling, further limiting their visibility.  (3) High-mass white 
dwarfs may be prone to be ejected from their parent clusters, either due to dynamical 
interactions or velocity kicks resulting from asymmetric mass loss during their formation (Fellhauer 
et~al.\ 2003; Tremblay et~al.\ 2012).

Our search for ultramassive white dwarfs begins with the very rich NGC 2099 with a large population 
of 50 white dwarf candidates (e.g., Kalirai et~al.\ 2001; 2005; Paper I).  
In Paper I we spectroscopically 
confirmed the white dwarf nature of 19 of the brighter white dwarf candidates in the cluster and 
measured their masses.  That work set the bulk of the constraints on the intermediate mass 
range of the IFMR (e.g., M$_{\rm initial}$ = 2.5 to 4.0~M$_\odot$).  In this letter, we 
push the initial study to fainter luminosities in search of more massive white dwarfs.

In Section 2 we discuss the spectroscopic white dwarf observations of NGC 2099 and describe 
the reduction and analysis techniques.  In Section 3 we discuss the cluster membership of the 
white dwarf candidates in NGC 2099.  In Section 4 we look at the M$_{\rm initial}$ and M$_{\rm final}$ of 
each white dwarf cluster member and analysis in detail the errors of ultramassive white dwarfs.  
In Section 5 we summarize our results.

\section{Observations, Reductions \& Analysis}

Our previous Keck I Low Resolution Imaging Spectrometer (LRIS; Oke et~al.\ 1995) observations of 
NGC 2099, presented in Paper I, observed a faint candidate (WD33) at $V$ = 24.49$\pm$0.065.  The 
resulting 
WD33 spectrum was not suitable for publication, but it suggested that this faint white dwarf had 
a high mass.  We obtained new Keck/LRIS observations during 2015 February 18 and 19 with a slitmask to 
re-observe WD33 and eight new white dwarf candidates in NGC 2099.  These additional eight targets 
span $V$ from 22.3 to 24.3 and were selected based on the 11 white dwarfs 
in Paper I that were found to be consistent with NGC 2099 membership.  Five hours of 
observation were acquired on this mask.  

Continuing with the methods from Paper I and II, we reduced and flux 
calibrated the new LRIS observations using the IDL based XIDL pipeline\footnote[7]{Available at 
http://www.ucolick.org/$\sim$xavier/IDL/}.  Of the total observed sample of nine white dwarf 
candidates, seven are DA white dwarfs, one is a DB white dwarf, and one is a DZ white dwarf.  
The new observations of WD33 have been coadded to the original observations taken
with Keck/LRIS under the same configuration.

For the spectroscopic DA analysis, we adopted the same techniques as described
in Paper II but with updated oxygen/neon (ONe) white dwarf models.  In brief, we used the 
white dwarf spectroscopic models of Tremblay et~al.\ 
(2011) with the Stark profiles of Tremblay \& Bergeron (2009), and the automated fitting techniques 
from Bergeron et~al.\ (1992) to fit the Balmer line spectra and derive T$_{\rm eff}$ 
and log g.  For the spectroscopic DB analysis, we adopted the methods in Bergeron et~al.\ (2011).  
For deriving M$_{\rm final}$, luminosity, and cooling age of the lower mass ($<$1.10 M$_\odot$) DA white dwarfs 
and the DB white dwarf, the cooling models by Fontaine et~al.\ (2001) were used for a carbon/oxygen 
(CO) core composition with a thick and thin hydrogen layer, for hydrogen and helium atmospheres, respectively.  
Lastly, for massive white dwarfs up to 1.28 M$_\odot$ we derived M$_{\rm final}$, luminosity, and cooling 
age based on the ONe-core models of Althaus et~al.\ (2007), up to 1.38 M$_\odot$ we used
unpublished ultramassive models using consistent physics (L.G. Althaus; private communication 2016).  
This both expands the mass range and updates our adopted ONe mass-radius relationship to 
that from the Althaus et~al.\ (2007) models.  In contrast, the Paper II analysis used the older 
mass-radius relationship from Althaus et~al.\ (2005).

\tablefontsize{\footnotesize}
\begin{deluxetable*}{l c c c c c c c c c c c c}
\multicolumn{13}{c}%
{{\bfseries \tablename\ \thetable{} - White Dwarf Initial \& Final Parameters}} \\
\hline \\
\\[-0.55cm]\hspace{-0.1cm}ID&\hspace{-0.1cm}M$_{\rm V}$       &\hspace{-0.1cm}B-V$_0$     &\hspace{-0.1cm}V &\hspace{-0.1cm}B-V  &\hspace{-0.1cm}$\alpha$&$\delta$               &\hspace{-0.1cm}T$_{\rm eff}$&\hspace{-0.1cm}log g&\hspace{-0.1cm}M$_{\rm final}$   &\hspace{-0.1cm}t$_{\rm cool}$&\hspace{-0.1cm} M$_{\rm initial}$     &\hspace{-0.1cm} S/N\\
\hspace{-0.1cm}  &\multicolumn{2}{c}{Model}& \multicolumn{2}{c}{Obs}                                            &\hspace{-0.1cm}(J2000) &\hspace{-0.1cm}(J2000) &\hspace{-0.1cm}(K)          &\hspace{-0.1cm}     &\hspace{-0.1cm}(M$_\odot$)    &\hspace{-0.1cm}(Myr)         &\hspace{-0.1cm}(M$_\odot$)&\hspace{-0.1cm}\\
\hline \\
\\[-0.55cm]\multicolumn{13}{l}{{Likely DA White Dwarf Members of NGC 2099}}  \\
\hline\\
\\[-0.55cm]
\hspace{-0.1cm}WD25  &\hspace{-0.1cm} 10.31 &\hspace{-0.1cm} -0.18 &\hspace{-0.1cm} 22.30 &\hspace{-0.1cm} 0.16 & \hspace{-0.1cm}05:52:44.44 & \hspace{-0.1cm}+32:29:54.7 & \hspace{-0.1cm}27500$\pm$450  & \hspace{-0.1cm}8.11$\pm$0.06 & \hspace{-0.1cm}0.70$\pm$0.03          & \hspace{-0.1cm} 17$^{+5}_{-3}$     & \hspace{-0.1cm}2.95$^{+0.01}_{-0.01}\,^{+0.10}_{-0.10}$ & \hspace{-0.1cm}82\\
\\[-0.25cm]\hspace{-0.1cm}WD28  &\hspace{-0.1cm} 10.89 &\hspace{-0.1cm} -0.09 &\hspace{-0.1cm} 22.73 &\hspace{-0.1cm} 0.20 & \hspace{-0.1cm}05:52:44.37 & \hspace{-0.1cm}+32:25:22.4 & \hspace{-0.1cm}22000$\pm$400  & \hspace{-0.1cm}8.20$\pm$0.06 & \hspace{-0.1cm}0.75$\pm$0.03          & \hspace{-0.1cm} 76$^{+13}_{-12}$   & \hspace{-0.1cm}3.07$^{+0.03}_{-0.03}\,^{+0.13}_{-0.11}$ & \hspace{-0.1cm}76\\
\\[-0.25cm]\hspace{-0.1cm}WD33  &\hspace{-0.1cm} 12.29 &\hspace{-0.1cm} -0.31 &\hspace{-0.1cm} 24.49 &\hspace{-0.1cm} 0.07 & \hspace{-0.1cm}05:52:36.35 & \hspace{-0.1cm}+32:27:16.8 & \hspace{-0.1cm}32900$\pm$1100 & \hspace{-0.1cm}9.27$\pm$0.22 & \hspace{-0.1cm}1.28$^{+0.05}_{-0.08}$ & \hspace{-0.1cm}233$^{+102}_{-118}$ & \hspace{-0.1cm}3.58$^{+0.62}_{-0.41}\,^{+0.25}_{-0.20}$ & \hspace{-0.1cm}22\\
\hline
\hline \\
\\[-0.55cm]\multicolumn{13}{l}{{DA White Dwarf Inconsistent with Single Star Membership of NGC 2099}}  \\
\hline\\
\\[-0.55cm]
\hspace{-0.1cm}WD29  &\hspace{-0.1cm} 11.41 &\hspace{-0.1cm} 0.01 &\hspace{-0.1cm} 23.13 &\hspace{-0.1cm}  0.41 &\hspace{-0.1cm} 05:53:04.82 &\hspace{-0.1cm} +32:29:26.0 &\hspace{-0.1cm} 17300$\pm$500  & \hspace{-0.1cm}8.26$\pm$0.10 & \hspace{-0.1cm}0.77$\pm$0.06 & \hspace{-0.1cm}195$^{+39}_{-34}$  & \hspace{-0.1cm}-- & \hspace{-0.1cm}36\\
\hline
\hline \\
\\[-0.55cm]\multicolumn{13}{l}{{DB White Dwarf in the field of NGC 2099}}  \\
\hline\\
\\[-0.55cm]
\hspace{-0.1cm}WD27  &\hspace{-0.1cm} 11.68 &\hspace{-0.1cm} -0.07 &\hspace{-0.1cm} 22.60 &\hspace{-0.1cm} 0.14 &\hspace{-0.1cm} 05:52:45.31 &\hspace{-0.1cm} +32:25:49.4 &\hspace{-0.1cm} 22100$\pm$120 &\hspace{-0.1cm} 8.66$\pm$0.07 & \hspace{-0.1cm}1.01$\pm$0.05 & \hspace{-0.1cm}204$^{+46}_{-40}$  & \hspace{-0.1cm}-- & \hspace{-0.1cm}67\\
\hline
\hline \\
\\[-0.55cm]\multicolumn{13}{l}{{Low Signal to Noise DA White Dwarfs and a DZ White Dwarf in the field of NGC 2099}}  \\
\hline\\
\\[-0.55cm]
\hspace{-0.1cm}WD30  &\hspace{-0.1cm} 11.26 &\hspace{-0.1cm} -0.02 &\hspace{-0.1cm} 23.66 &\hspace{-0.1cm} 0.28 &\hspace{-0.1cm} 05:53:03.06 &\hspace{-0.1cm} +32:26:12.4 &\hspace{-0.1cm} 18200$\pm$950  &\hspace{-0.1cm} 8.22$\pm$0.16 &\hspace{-0.1cm} 0.75$\pm$0.11 &\hspace{-0.1cm} 158$^{+63}_{-50}$ &\hspace{-0.1cm} -- &\hspace{-0.1cm}  18\\
\\[-0.25cm]\hspace{-0.1cm}WD31  &\hspace{-0.1cm} 11.06 &\hspace{-0.1cm}  0.06 &\hspace{-0.1cm} 24.26 &\hspace{-0.1cm} 0.35 &\hspace{-0.1cm} 05:52:53.69 &\hspace{-0.1cm} +32:30:11.3 &\hspace{-0.1cm} 14400$\pm$1200 &\hspace{-0.1cm} 7.80$\pm$0.21 &\hspace{-0.1cm} 0.50$\pm$0.11 &\hspace{-0.1cm} 165$^{+82}_{-57}$ &\hspace{-0.1cm} -- &\hspace{-0.1cm}  20\\
\\[-0.25cm]\hspace{-0.1cm}WD32  &\hspace{-0.1cm} 11.04 &\hspace{-0.1cm} -0.09 &\hspace{-0.1cm} 24.34 &\hspace{-0.1cm} 0.26 &\hspace{-0.1cm} 05:53:01.44 &\hspace{-0.1cm} +32:26:42.0 &\hspace{-0.1cm} 22400$\pm$2000 &\hspace{-0.1cm} 8.31$\pm$0.28 &\hspace{-0.1cm} 0.82$\pm$0.17 &\hspace{-0.1cm}  94$^{+82}_{-57}$ &\hspace{-0.1cm} -- &\hspace{-0.1cm}  12\\
\\[-0.25cm]\hspace{-0.1cm}WD26  &\hspace{-0.1cm}  --   &\hspace{-0.1cm}   --  &\hspace{-0.1cm} 22.44 &\hspace{-0.1cm} 0.36 &\hspace{-0.1cm} 05:53:07.18 &\hspace{-0.1cm} +32:28:59.9 &\hspace{-0.1cm}     --         &\hspace{-0.1cm}      --       &\hspace{-0.1cm}     --        &\hspace{-0.1cm}        --         &\hspace{-0.1cm} -- &\hspace{-0.1cm}  50\\
\hline
\hline \\
\\[-0.55cm]\multicolumn{13}{l}{{Massive white dwarf members of NGC 3532 and the Pleiades.}}  \\
\hline\\
\\[-0.55cm]
\multicolumn{5}{l}{VPHASJ1103-5837} &\hspace{-0.1cm} 11:03:58.00 & \hspace{-0.1cm}-58:37:09.2 & \hspace{-0.1cm}23900$\pm$450  & \hspace{-0.1cm}8.87$\pm$0.06 & \hspace{-0.1cm}1.11$\pm$0.03 & \hspace{-0.1cm}223$^{+40}_{-30}$  & \hspace{-0.1cm}5.40$^{+1.36}_{-0.55}$ & \hspace{-0.1cm}35\\
\\[-0.25cm]\multicolumn{5}{l}{GD50           } &\hspace{-0.1cm} 03:46:17.26 & \hspace{-0.1cm}-01:07:31.5 & \hspace{-0.1cm}42700$\pm$800  & \hspace{-0.1cm}9.20$\pm$0.07 & \hspace{-0.1cm}1.26$\pm$0.02 & \hspace{-0.1cm} 76$^{+17}_{-11}$  & \hspace{-0.1cm}6.41$^{+0.72}_{-0.41}$ & \hspace{-0.1cm}--\\
\\[-0.25cm]\multicolumn{5}{l}{PG 0136+251    } &\hspace{-0.1cm} 01:38:53.02 & \hspace{-0.1cm}+25:23:22.8 & \hspace{-0.1cm}41400$\pm$800  & \hspace{-0.1cm}9.03$\pm$0.07 & \hspace{-0.1cm}1.20$\pm$0.03 & \hspace{-0.1cm} 52$^{+14}_{-12}$  & \hspace{-0.1cm}5.78$^{+0.48}_{-0.32}$ & \hspace{-0.1cm}--\\
\hline \\
\\[-0.55cm]\caption{The first M$_{\rm initial}$ errors are based on the white dwarf parameter errors and for NGC 2099 members the second M$_{\rm initial}$ errors are based on cluster
age errors.}
\end{deluxetable*}  

Table 1 presents the observed and derived parameters for the new white dwarf candidates from NGC 2099.  
We have organized these white dwarfs by type and membership (see Section 3), but we also separate 
the DZ WD26 because we cannot analyze it and WD30, WD31, and WD32 because they have very 
low S/N spectra with mass uncertainties $>$0.1~M$_\odot$.  Their membership analysis is 
unreliable so we did not use them in the IFMR.

\begin{figure}[!ht]
\begin{center}
\vspace{-0.2cm}
\includegraphics[clip, scale=0.8]{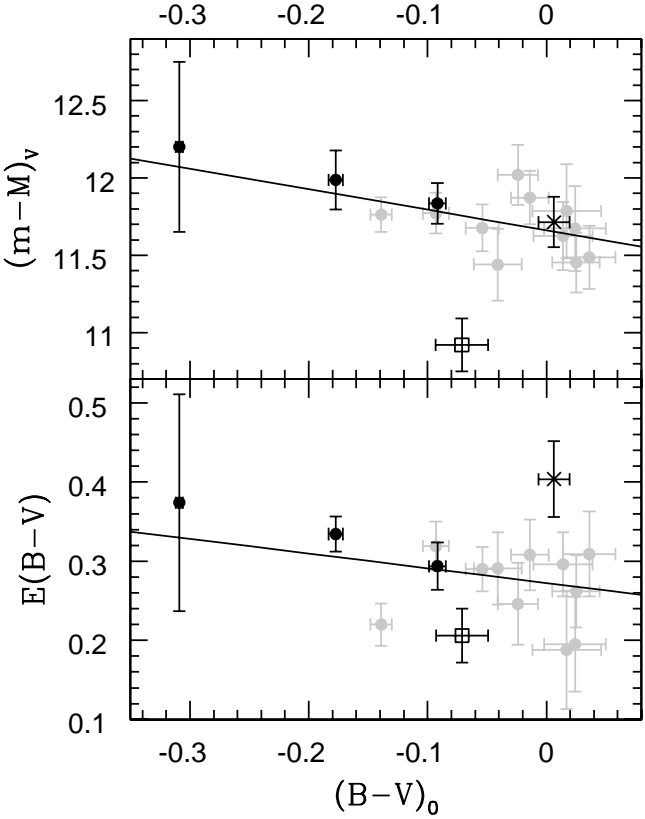}
\end{center}
\vspace{-0.4cm}
\caption{The upper panel shows the effective distance modulus for the DA members (solid black) and 
nonmember (x), and the DB white dwarf (open square).  The data are plotted versus predicted 
(B-V)$_0$ and are compared to the NGC 2099 members from Paper I (solid gray).  The lower panel 
shows the effective reddening versus predicted (B-V)$_0$.  The solid lines illustrate
the color trends for distance modulus and reddening.  All white dwarfs are plotted
with their 1$\sigma$ error bars, and white dwarfs within 2$\sigma$ of the trend in both distance 
modulus and reddening are considered members.}
\end{figure}

Table 1 also includes the newly discovered VPHASJ1103-5837 from NGC 3532 (Raddi et~al.\ 2016)
and updated initial and final-masses for GD50 and PG 0136+251 (Gianninas et~al.\ 2011).  The 
spectroscopic analysis techniques in both studies were equivalent to ours, so we applied their 
T$_{\rm eff}$ and log g directly (we added external errors [see Paper I] to 
VPHASJ1103-5837's published errors), and we derived both the masses and cooling ages from the ONe 
models of Althaus et~al.\ (2007).  

\section{White Dwarf Membership in NGC 2099}

To apply these white dwarfs to the IFMR, cluster membership must be verified to
be able to infer their M$_{\rm initial}$.  For WD33, the significant
mass and high T$_{\rm eff}$ is by itself a strong argument for membership, but to
refine its membership and determine the membership status of the other observed white dwarfs
we compared the predicted colors and magnitudes to the photometry (see Table 1).  
This is similar to the procedure from Paper I, where we compared to the 
NGC 2099 photometry from Kalirai et~al.\ (2001), but we now have an expanded sample 
and color range to both refine the white dwarf based distance modulus and reddening and to also 
look for trends with color.  

Figure 1 compares the apparent distance moduli and reddenings for each observed
white dwarf with sufficient signal and plots them versus their model-based predicted color.
Their 1$\sigma$ distance modulus and reddening errors are shown, which are the photometric and 
model-based errors added in quadrature.  In both cases, we find color trends for distance modulus 
and reddening.  These trends may be the result of the photometric standardization, 
\begin{figure}[!ht]
\begin{center}
\vspace{-0.2cm}
\includegraphics[clip, scale=0.6]{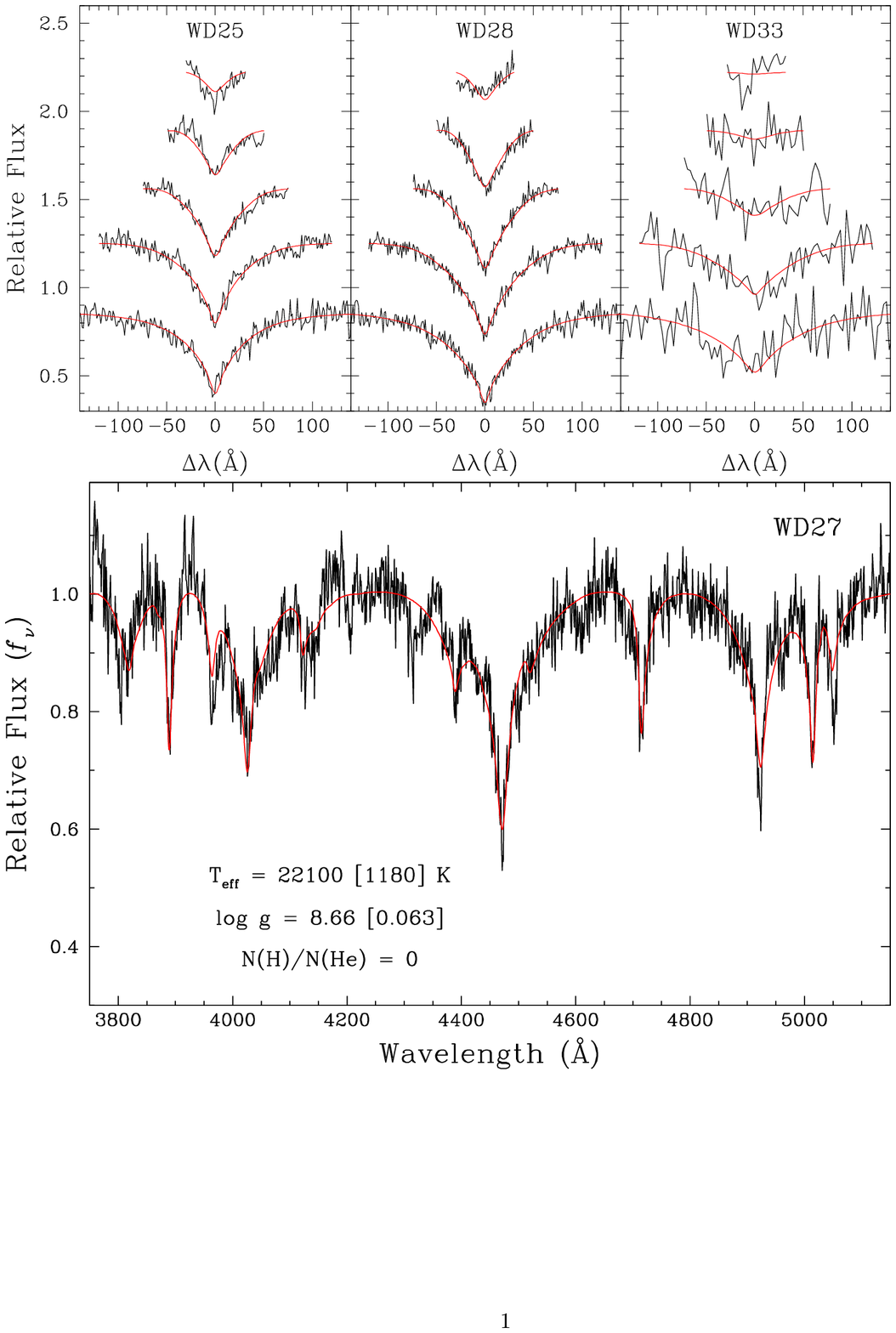}
\end{center}
\vspace{-0.4cm}
\caption{The upper three panels show the Balmer line fits for the three white dwarf members 
of NGC 2099.  The spectrum of WD33 has been binned for display purposes.  The H$\beta$, 
H$\gamma$, H$\delta$, H$\epsilon$, and H8 fits are shown from bottom to top.  The lower panel 
shows the fit of WD27's He features, where we have adopted a pure He atmosphere.}
\end{figure}
which can be less precise in both blue stars and in faint stars.  Additionally, the reddening in
NGC 2099 is quite large, and as discussed in Paper I (see also Fernie et~al.\ 1963) at reddenings 
E(B-V)$>$0.2 the effective reddening and extinction are meaningfully dependent on intrinsic color.
We find WD25, WD28, and WD33 are consistent with single star membership in NGC 2099 because 
they are within 2$\sigma$ of the trend in both distance modulus and reddening.

\begin{figure*}[!ht]
\begin{center}
\vspace{-0.1cm}
\includegraphics[clip, scale=0.92]{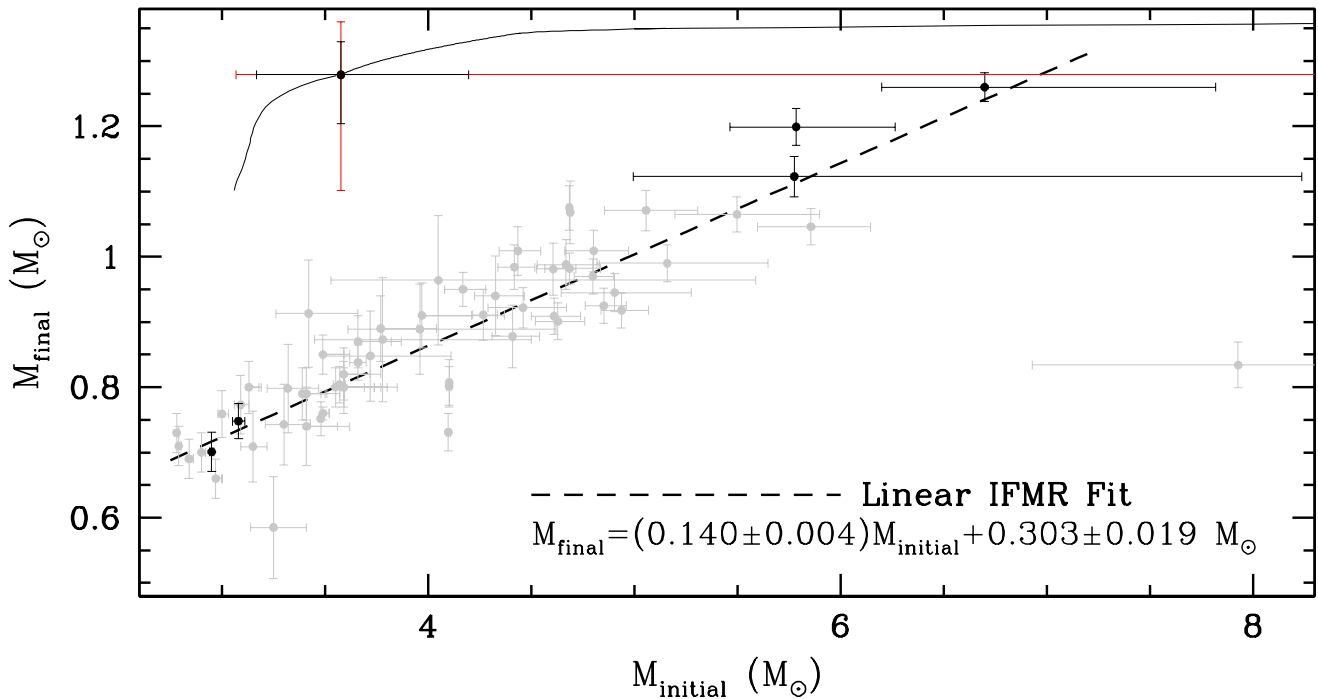}
\end{center}
\vspace{-0.4cm}
\caption{The IFMR data from Paper I and II (gray) are plotted with the three newly observed members
of NGC 2099 (black).  WD25 and WD28 are strongly consistent with the previous data, while the ultramassive
WD33 gives a very low M$_{\rm initial}$ but with significant mass errors (1$\sigma$-black; 2$\sigma$-red).  
Because the initial and
final mass errors in WD33 are not independent, we also display a curve showing the direct and strong
relation between adopted M$_{\rm final}$ and the resulting M$_{\rm initial}$.  We also include
VPHASJ1103-5837 and the updated Pleiades white dwarfs.  The fit relation displayed
does not consider WD33 due to its significant errors.}
\end{figure*}

The observed photometry of the DB white dwarf WD27 is 0.84$\pm$0.17 magnitudes too bright to be 
consistent with single star membership.  However, if it is a binary member of 
two comparable luminosity white dwarfs its observed magnitude would be 
$\sim$0.75 magnitudes brighter than the model predicts.  The inferred reddening of this DB white 
dwarf is $\sim$2$\sigma$ lower than expected for a member, which may suggest it is a less-reddened 
foreground DB white dwarf, but it is 
still within the reddening membership criterion.  Additionally, its younger cooling age of 204 Myr is 
well within the NGC 2099 cluster age of 520$\pm$50 Myr (Kalirai et~al.\ 2001; Paper I).  However,
we note the unlikelihood of a binary with two nearly equivalent DB white dwarfs and the lack of 
Balmer features in the spectrum (see Figure 2) that would indicate a DA companion.  Irrespective
of membership, WD27 is an interesting and very rare DB because it is both moderately hot and high mass 
(see Bergeron et~al.\ 2011; Koester \& Kepler 2015).

Figure 2 displays the spectral fits of the three white dwarf members WD25, WD28, and WD33, and
the DB white dwarf WD27.  While the WD33 spectrum has low S/N, most notably at the two 
highest-order Balmer lines, at this high mass and moderate T$_{\rm eff}$ these highest-order 
lines become increasingly less sensitive to log g.  For example, fitting only the first four 
Balmer lines derives log g=9.23$\pm$0.22 and only the first three lines derives log 
g=9.30$\pm$0.24.  Lastly, spectral analysis of 831 synthetic spectra with input 
parameters of T$_{\rm eff}$=32,900 K and log g=9.27 and S/N=22 finds a normally distributed series 
of log g measurements with a mean consistent with the input, and the distribution's $\sigma$ 
matches our spectral analysis's fitting error.

\vspace{0.6cm}
\section{Initial-Final Mass Relation}

We measured the IFMR by comparing each white dwarf's cooling age to the NGC 2099 cluster age 
(520 Myr).  
The difference between these ages gives the evolutionary time to the tip of the AGB 
for each white dwarf's progenitor.  We applied these times to the PARSEC evolutionary 
models (Bressan et~al.\ 2012) to determine each white dwarf's M$_{\rm initial}$.  
These M$_{\rm initial}$ values are given in Table 1, including two M$_{\rm initial}$ errors based
on the white dwarf parameter errors and from the cluster 
age errors (520$\pm$50 Myr).  For the M$_{\rm initial}$ of VPHASJ1103-5837 we adopted for NGC 3532
the Paper II cluster age of 320 Myr.  For GD50 and PG 0136+251
we adopted for Pleiades the Paper II cluster age of 135 Myr.

\begin{figure}[!ht]
\begin{center}
\vspace{-0.1cm}
\includegraphics[clip, scale=0.78]{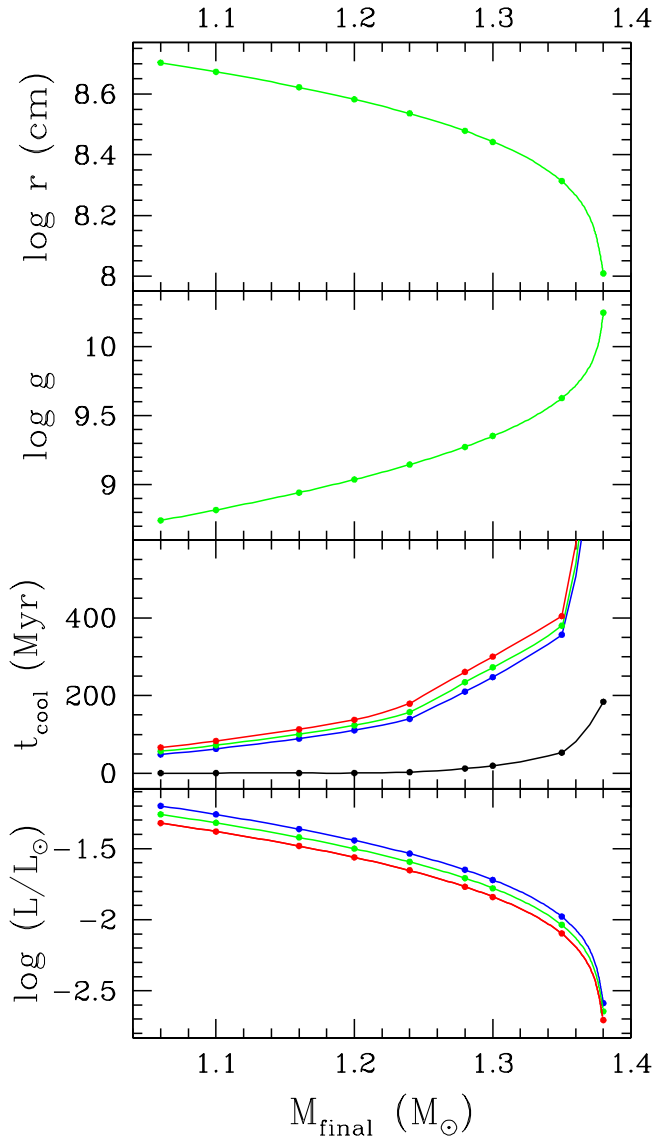}
\end{center}
\vspace{-0.4cm}
\caption{These panels illustrate how log radius, log g, cooling age, and log L/L$_\odot$ vary with
M$_{\rm final}$ at constant T$_{\rm eff}$ in the ONe models of Althaus et~al.\ (2007) plus
consistent higher-mass models.  Green 
represents the derived T$_{\rm eff}$ of WD33 at 32,900 K.  The lower two panels illustrate
the effects of the $\pm$1100 K T$_{\rm eff}$ errors by plotting 34,000 K (blue) and 31,800 K (red).
The cooling age at 1.38 M$_\odot$ is not displayed but is 1168$^{+114}_{-101}$ Myr.
For cooling age, we also illustrate the weakened sensitivity to mass at a higher T$_{\rm eff}$
of 65,000 K (black).}
\end{figure}

Beginning with the high-mass DB white dwarf WD27, if WD27 is a double degenerate consistent with
cluster membership it may have undergone some level
of binary interaction in its past.  This could potentially explain its nature, but this would also 
make its inferred M$_{\rm initial}$ unreliable.  Testing this, its relatively short cooling time of 
204$^{+39}_{-34}$ Myr implies a M$_{\rm initial}$ of only 3.46$^{+0.16}_{-0.12}$ M$_\odot$, 
while our IFMR fit gives that a singly evolved 1.0 M$_\odot$ white dwarf would have a 
$\sim$5.0 M$_\odot$ progenitor.  

Figure 3 compares the Paper I and II IMFR data with VPHASJ1103-5837, the updated Pleiades white 
dwarfs, and the three newly observed NGC 2099 members.  WD25, WD28, VPHASJ1103-5837, PG 0136+251, 
and GD50 are strongly consistent with the Paper II IFMR trend.  The ultramassive WD33, however, is 
very discrepant, but there are
several possible explanations for this.  First, is it a supermassive white dwarf formed through a 
merger of two lower-mass white dwarfs?  Based on the models of white dwarf mergers from Dan et~al.\ 
(2014), the mechanism to create a stable supermassive white dwarf requires the merging of two 
comparable white dwarfs of approximately half its mass ($\sim$0.64 M$_\odot$).  Based on the age of 
NGC 2099 this is pushing the minimum mass of a white dwarf that could have formed after 520 Myr.  
Binary interactions could have affected their evolutionary timescales, but they still would 
have likely just formed in the recent past and would not have had the time to both merge to
create WD33 and subsequently cool for 233 Myr.  

Second, WD33 could be the result of a binary merger event that occurred while the components 
were still evolving.  Two binary components of both $\sim$3.5 M$_\odot$ could have undergone 
interaction and subsequent merger.  This would have created a short-lived $\sim$7 M$_\odot$ 
blue straggler that quickly formed into WD33 and had sufficient time to still cool for 233 Myr.  

Lastly, another possibility relates to both the analysis and systematic errors in the ONe 
cooling models themselves.  The errors in black in Figure 3 are the 1$\sigma$ errors in both 
M$_{\rm final}$ and M$_{\rm initial}$.  Expanding our error analysis in WD33 to look at 2$\sigma$ 
variations in log g (in red) finds that at higher masses the uncertainty in the cooling age rapidly 
expands.  This results from a white dwarf's radius becoming increasingly sensitive to
mass in this regime (Althaus et~al.\ 2005; 2007).  Figure 4 displays the Althaus et~al.\ (2007) 
mass-radius relationship at WD33's T$_{\rm eff}$ of 32,900 K.  This mass sensitivity 
in radius also leads to a significant sensitivity in log g, cooling age, and luminosity at higher 
masses.  In 
Figure 4, we also analyze the sensitivity to WD33's 1$\sigma$ T$_{\rm eff}$ error (1100 K) for 
cooling age and luminosity, with a high-T$_{\rm eff}$ (blue) and low-T$_{\rm eff}$ (red) curve.  
This illustrates that at the highest masses the sensitivity of the cooling age to T$_{\rm eff}$ errors 
is minor relative to the mass dependence.  

In application to the ultramassive IFMR, one advantage of these strong dependencies is that 
large errors in the spectroscopically derived log g result in only moderate to minor errors 
in M$_{\rm final}$.  A second advantage is that the increasing sensitivity of luminosity 
to M$_{\rm final}$ can be used to independently infer mass from photometry, but 
uncertainties in WD33's observed magnitude and NGC 2099's visual distance modulus currently limit 
how accurately we can observationally derive its M$_{\rm V}$.  

The significant challenge for the ultramassive IFMR is the extreme sensitivity to M$_{\rm final}$ 
of cooling age, and hence M$_{\rm initial}$.  As seen in Table 1 and Figure 4, WD33's
parameters only derive a modest cooling age of 233 Myr, and with the adopted cluster age of 520 Myr this
gives a very low M$_{\rm final}$ of 3.58 M$_\odot$.  A 1$\sigma$ increase in WD33's mass to 
1.33 M$_\odot$ increases the derived cooling age to 331 Myr.  While a 2$\sigma$ increase in white dwarf 
mass, from propagating a 2$\sigma$ increase in log g, to 1.36 M$_\odot$ increases the cooling age to 
546 Myr, surpassing the cluster age.  Figure 3 demonstrates this strong dependence of initial
and final mass errors with a single curve passing through WD33.  This also illustrates the additional
challenge that as M$_{\rm initial}$ increases the sensitivity of derived M$_{\rm initial}$ to evolutionary 
lifetime increases rapidly.  

These cooling age challenges, reassuringly, do not equally affect all ultramassive white dwarfs.  
The youngest and hottest white dwarfs in this mass range are significantly less susceptible to 
these complications (e.g., GD50).  First, these young white dwarfs are higher luminosity, increasing
the ease of acquiring high-signal spectra.  Second, the sensitivity of cooling age on white dwarf
mass significantly decreases at high T$_{\rm eff}$.  For example, at WD33's T$_{\rm eff}$ of 32,900 
K, a decrease in 
M$_{\rm final}$ from 3.36 to 3.34 M$_\odot$ causes a 191 Myr decrease in inferred cooling age.  
For a young white dwarf at T$_{\rm eff}$ of 65,000 K, this same change in M$_{\rm final}$ would 
result in a decrease in inferred cooling age of 37 Myr (see full comparison in Figure 4).  
A third advantage for young and ultramassive white dwarfs is that cooling ages are 
further complicated by dependencies on both the input physics and composition 
in the adopted cooling model, where potential systematics introduced in the cooling age grow
rapidly with increasing cooling age.  

\vspace{0.2cm}
\section{Summary}

We have observed nine new white dwarf candidates in NGC 2099.  Two intermediate-mass (WD25, WD28)
and one ultramassive (WD33) DA white dwarfs were found to be consistent with membership.
We also compared to the self-consistently analyzed GD50, PG 0136+251, and the newly discovered 
VPHASJ1103-5837.  Application of these data to the IFMR finds strong consistency with our
previous work for all but WD33, but this may be explained by WD33's significant M$_{\rm initial}$
errors.  Acquiring additional spectroscopic signal on WD33 may be of interest, and more 
accurate photometry would also be useful, but overcoming these errors at this mass and T$_{\rm eff}$ 
currently may not be viable at V=24.49.  For precise application of ultramassive white dwarfs to 
the IFMR, future studies should focus on clusters of age $\sim$50 to 150 Myr.  Nevertheless, because 
GD50 and PG 0136+251 are only kinematically connected to the Pleiades, WD33 is the first ultramassive 
white dwarf that is photometrically consistent with membership in a star cluster.  

\vspace{0.7cm}
This project was supported by the National Science Foundation (NSF) through grant AST-1211719.
This work was also supported by a NASA Keck PI Data Award, administered by the NASA Exoplanet
Science Institute. Data presented herein were obtained at the WM Keck Observatory from
telescope time allocated to the National Aeronautics and Space Administration through the agency's
scientific partnership with the California Institute of Technology and the University of California.
The Observatory was made possible by the generous financial support of the WM Keck Foundation.

\end{document}